\documentclass{article}

\usepackage[section]{placeins}
\usepackage{tabularx}
\usepackage{authblk}
\usepackage{subcaption}
\usepackage{graphicx}
\usepackage{hyperref}
\usepackage{amsmath}
\usepackage{mathtools}

\begin{document}

\title{Quantifying polaronic effects on charge-carrier scattering and mobility in lead--halide perovskites}
\author{Matthew J. Wolf}
\author{Lewis A. D. Irvine}
\author{Alison B. Walker}
\affil{Department of Physics, University of Bath, Claverton Down, Bath, BA2 7AY, United Kingdom}
\date{}

\maketitle

\begin{abstract}
The formation of polarons due to the interaction between charge carriers and the crystal lattice has been proposed to have wide-ranging effects on charge carrier dynamics in lead--halide perovskites (LHPs). The hypothesis underlying many of those proposals is that charge carriers are ``protected'' from scattering by their incorporation into polarons. We test that hypothesis by deriving expressions for the rates of scattering of polarons by polar-optical and acoustic phonons, and ionised impurities, which we compute for electrons in the LHPs MAPbI$_{3}$,  MAPbBr$_{3}$ and CsPbI$_{3}$. We then use the ensemble Monte Carlo method to compute electron-polaron distribution functions which satisfy a Boltzmann equation incorporating the same three scattering mechanisms. By carrying out analogous calculations for band electrons and comparing their results to those for polarons, we conclude that polaron formation impacts charge-carrier scattering rates and mobilities to a limited degree in LHPs, contrary to claims in the recent literature.
\end{abstract}

\pagebreak


Lead--halide perovskites (LHPs) are currently the subject of an intense international research effort, primarily due to their application as an active layer in next generation optoelectronic devices, and photovoltaic cells in particular. However, there remain a number of fundamental aspects of their optoelectronic properties which remain under debate, such as the origins of the observed long carrier lifetimes, the seemingly benign nature of defects, and the mobility limiting scattering mechanism(s) \cite{egger2018remains, herz2018lattice}.

In materials with polar bonding, such as LHPs, the electrostatic interaction between a charge carrier and the lattice ions in its vicinity causes the latter to be displaced from their equilibrium positions. The charge carrier, along with the polarised region of the lattice which surrounds it, comprises a quasi-particle known as a polaron. A distinction is typically made between the self-trapped ``small'' polaron, and the itinerant ``large'' polaron; in the remainder of this paper, we use the general term ``polaron'' to refer to the latter species exclusively. Polaron formation has been suggested to play a central role in numerous elementary processes underlying charge carrier dynamics in LHPs \cite{zhu2015charge, miyata2017lead, wolf2017polaronic}, including exciton dissociation \cite{soufiani2015polaronic, menendez2015nonhydrogenic, bokdam2016role, straus2016direct}, hot carrier cooling \cite{niesner2016persistent, zhu2016screening, frost2017slow, guo2017long, evans2018competition, joshi2019dynamic}, radiative and non-radiative recombination \cite{chen2016extended, ivanovska2017long, ambrosio2018origin, he2018photoinduced, munson2018dynamic, wiktor2018mechanism, munson2019vibrational} and steady state mobilities \cite{chen2016extended, sendner2016optical, mante2017electron}.

The most widely recognised consequence of polaron formation is an increase in the effective mass of a charge carrier from that predicted by band theory. In LHPs, the increase has been calculated to be of the order of 30--70{\%} \cite{sendner2016optical, frost2017calculating, schlipf2018carrier, puppin2019evidence} at room temperature. However, the hypothesis underlying many of the proposals of ways in which polaron formation influences charge carrier dynamics, is that charge carriers are ``protected'' from interactions with lattice vibrations, defects and other charge carriers by their incorporation into polarons \cite{zhu2015charge}. 

The aim of this paper is to provide a quantitative analysis of that hypothesis, building primarily on the seminal work on polaron theory of R. P. Feynman \cite{feynman1955slow}, Y. Osaka \cite{osaka1959polaron} and L. P. Kadanoff \cite{kadanoff1963boltzmann}. We derive angular dependent rates of scattering of polarons by acoustic phonons and ionised impurities, and compute their values, along with those for scattering by polar-optical phonons, for electron polarons in the LHPs MAPbI$_{3}$, MAPbBr$_{3}$ and CsPbI$_{3}$ (where MA represents methylammonium, CH$_3$NH$_3$). We then use those rates, along with the computed electron-polaron masses, in an augmented form of Kadanoff's semi-classical Boltzmann equation, from which we compute the steady state electron mobility using the ensemble Monte Carlo method. Comparing the results of those computations with their analogues for band electrons, we conclude that polaron formation has a much less significant impact than has been asserted in the recent literature. Due to the similarity between the results for the three materials, data for MAPbBr$_{3}$ and CsPbI$_{3}$ are given in the supplementary information. 

There have been a number of recent theoretical studies of polarons in LHPs based on density-functional-theory (DFT) calculations \cite{bischak2017origin, kang2017preferential, kang2017shallow, ambrosio2018origin, ambrosio2019charge, meggiolaro2019polarons, cinquanta2019ultrafast, zheng2019large}. The starting point of our analysis is instead the model introduced by R. P. Feynman \cite{feynman1955slow} in which the ``cloud of virtual phonons'' associated with the polarised region of the lattice surrounding the electron is represented by a fictitious particle, which is coupled to the electron via an harmonic potential. The corresponding Hamiltonian is \cite{peeters1984theory}:

\begin{equation}\label{eqn:hamiltonian1}
    H_{\mathrm{F}}
    =\frac{\hbar \lvert \mathbf{k} \rvert^{2}}{2m} + \frac{\hbar \lvert \mathbf{k}_{\mathrm{c}} \rvert^{2}}{2m_{\mathrm{c}}} + \frac{1}{2}\kappa \left (\mathbf{r}-\mathbf{r}_{\mathrm{c}} \right)^{2},
\end{equation} 

where $\mathbf{r}$, $\mathbf{k}$ and $m$ are the position, wave vector and effective mass of the electron, with analogous quantities for the phonon cloud being identified with a ``c'' subscript, and $\kappa$ is the spring constant of the harmonic potential.

The apparent simplicity of the Feynman model notwithstanding, it contains essential physics which DFT based methods do not, despite recent developments \cite{sio2019ab, sio2019polarons}. This is due to the Born--Oppenheimer approximation underlying those methods, in which the ions---the quantised vibrations of which constitute phonons---are treated as \emph{classical} particles, whereas the electron and the phonon cloud are treated on an equal \emph{quantum mechanical} footing in Eqn. \ref{eqn:hamiltonian1}. 

Eqn. \ref{eqn:hamiltonian1} may be re-written in terms of the centre-of-mass and relative co-ordinates of the two particle system as follows:

\begin{equation}\label{eqn:hamiltonian2}
    H_{\mathrm{F}}
    =\frac{\hbar \lvert \mathbf{K} \rvert^{2}}{2M}+\hbar\omega_{\mathrm{osc}}\sum_{i=1}^{3}\left(a_{i}^{\dagger}a_{i}+\tfrac{1}{2}\right), 
\end{equation} 

where $M$ and $\mathbf{K}$ are the total mass and wave vector of the polaron; $a_{i}^{\dagger}$ and $a_{i}$ are the ladder operators for the polaron's internal harmonic oscillator state, with the index $i$ labelling the three Cartesian directions; and $\omega_{\mathrm{osc}}$ is the angular frequency of the harmonic potential.

The eigenfunctions of Eqn. \ref{eqn:hamiltonian2} are of the form of a plane wave in the centre-of-mass co-ordinates, multiplied by a three dimensional harmonic oscillator state in the relative co-ordinates, and thus describe delocalised, itinerant states of a composite particle. We use the notation $\lvert \mathbf{K}, \mathbf{n}\rangle$ for the eigenstates, where  $\mathbf{K}$ is the polaron wave vector and  $\mathbf{n}=(n_{x}, n_{y}, n_{z})$ labels the polaron's internal oscillator state. 


Eqn. \ref{eqn:hamiltonian2} contains two free parameters, namely the polaron mass $M$ and the oscillator frequency $\omega_{\mathrm{osc}}$. Following previous theoretical studies on polarons in LHPs \cite{sendner2016optical, frost2017calculating}, we determine their values as functions of the lattice temperature by minimising the expression for the free energy derived by Y. Osaka \cite{osaka1959polaron}. The derivation of that expression assumes the presence of a single optical phonon branch, while the real phonon band structure possesses many. In order to circumvent that problem, we follow the approach of a previous study \cite{frost2017calculating} and use a single effective optical phonon angular frequency derived from the full optical phonon spectrum via the ``B'' scheme of Hellwarth and Biaggio \cite{hellwarth1999mobility}, which we label $\omega_{0}$.

\begin{figure}[h!]
    \centering
    \includegraphics[width=0.5\linewidth]{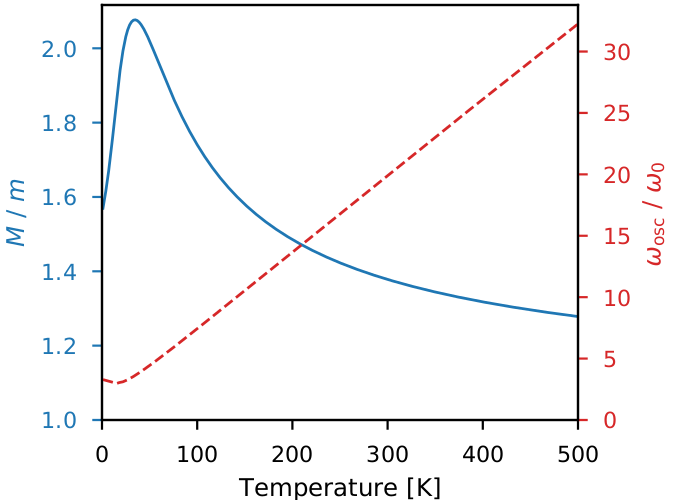}
    \caption{Temperature dependence of the mass $M$ (in units of $m$, the band electron effective mass) and oscillator frequency $\omega_{\mathrm{osc}}$ (in units of $\omega_{0}$, the effective polar optical phonon frequency) characterising the electron polaron in MAPbI$_{3}$. Similar results for MAPbBr$_{3}$ and CsPbI$_{3}$ are presented in the supplementary information.}
    \label{fig:polaron_properties}
\end{figure}

The results for electron polarons in MAPbI$_{3}$ are presented in Fig. \ref{fig:polaron_properties}; as expected, the calculated values are identical to those presented in Ref. \cite{frost2017calculating}.

The temperature dependent mass of the polaron, $M$, has a maximum value of $\sim2.1$ times the band-electron effective mass $m$, which occurs at $\sim35$ K. However, it decreases monotonically for temperatures above that value to the extent that at 300 K, it is only $\sim1.4 m$. We also note that the angular frequency $\omega_{\mathrm{osc}}$ (and therefore the excitation energy $\hbar \omega_{\mathrm{osc}}$) of the polaron harmonic oscillator, increases quasi-linearly with temperature above $\sim15$ K, which means that the equilibrium probability of a polaron being in its internal ground state $\mathbf{n}=\mathbf{0}$ is essentially independent of temperature. We plot the probability for MAPbI$_{3}$ (in addition to those for MAPbBr$_{3}$ and CsPbI$_{3}$) in the supplementary information; we simply note here that its value remains greater than 99.7 \% over the temperature range $0-500$ K. This has significant implications for the scattering rates of the polaron, which we now go on to address.

We follow Kadanoff in calculating the scattering rates for polarons using Fermi's golden rule, with eigenstates of Eqn. \ref{eqn:hamiltonian2}, $\lvert \mathbf{K}_{\mathrm{i}}, \mathbf{n}_{\mathrm{i}} \rangle$ and $\lvert \mathbf{K}_{\mathrm{f}}, \mathbf{n}_{\mathrm{f}} \rangle$, as the initial and final states \cite{kadanoff1963boltzmann}. We restrict ourselves to the case in which the polaron is in its internal ground state both before and after the scattering event, which we expect to be a valid approximation at, and close to, equilibrium in light of the  discussion in the previous paragraph.

The scattering rates are therefore calculated according to the following general expression:

\begin{equation}\label{eqn:goldenrule}
    S(\mathbf{K}_{\mathrm{i}} \rightarrow \mathbf{K}_{\mathrm{f}}) = \frac{2 \pi}{\hbar} \vert \langle \mathbf{K}_{\mathrm{f}}, \mathbf{0} \vert H_{\mathrm{pert}} \vert \mathbf{K}_{\mathrm{i}}, \mathbf{0} \rangle \rvert^{2} \delta(E_{\mathbf{K}_{\mathrm{f}}, \mathbf{0}} - E_{\mathbf{K}_{\mathrm{i}},\mathbf{0}}-\Delta E)
\end{equation}

In Eqn. \ref{eqn:goldenrule}, $H_{\mathrm{pert}}$ is a time-dependent perturbing Hamiltonian, the forms of which are well known for the main scattering mechanisms in polar semiconductors, namely acoustic and polar optical phonons, and ionised impurities \cite{jacoboni1989monte, tomizawa1993numerical}. The magnitude of $\Delta E$ in Eqn. \ref{eqn:goldenrule} depends on the scattering mechanism; for (quasi-)elastic scattering, such as that due to acoustic phonons and ionised impurties, $\Delta E=0$, while for polar optical phonon scattering, $\Delta E=\hbar \omega_{0}$.

The derivations of the scattering rates are provided in the supplementary information, in addition to those of the corresponding rates for band electrons. In summary, the differences between the scattering rates for polarons and those for band electrons amount to replacing $\mathbf{k}$ in the band-electron expressions with $\mathbf{K}$, and then multiplying by a factor $\exp{\left(-\lvert \mathbf{\mathbf{K}_{\mathrm{f}} - \mathbf{K}_{\mathrm{i}}} \rvert^{2} \hbar \mu / (2m^{2} \omega_{\mathrm{osc}}) \right)}$, where $\mu$ is the reduced mass of the polaron. As previously noted for the case of polar optical phonon scattering \cite{kadanoff1963boltzmann, peeters1984theory}, the exponential factor should suppress large changes in wave vector, ostensibly supporting the hypothesis that polarons are protected from scattering in comparison to their band carrier counterparts.

\begin{figure}[t!]
    \centering
    \includegraphics[width=1.0\linewidth]{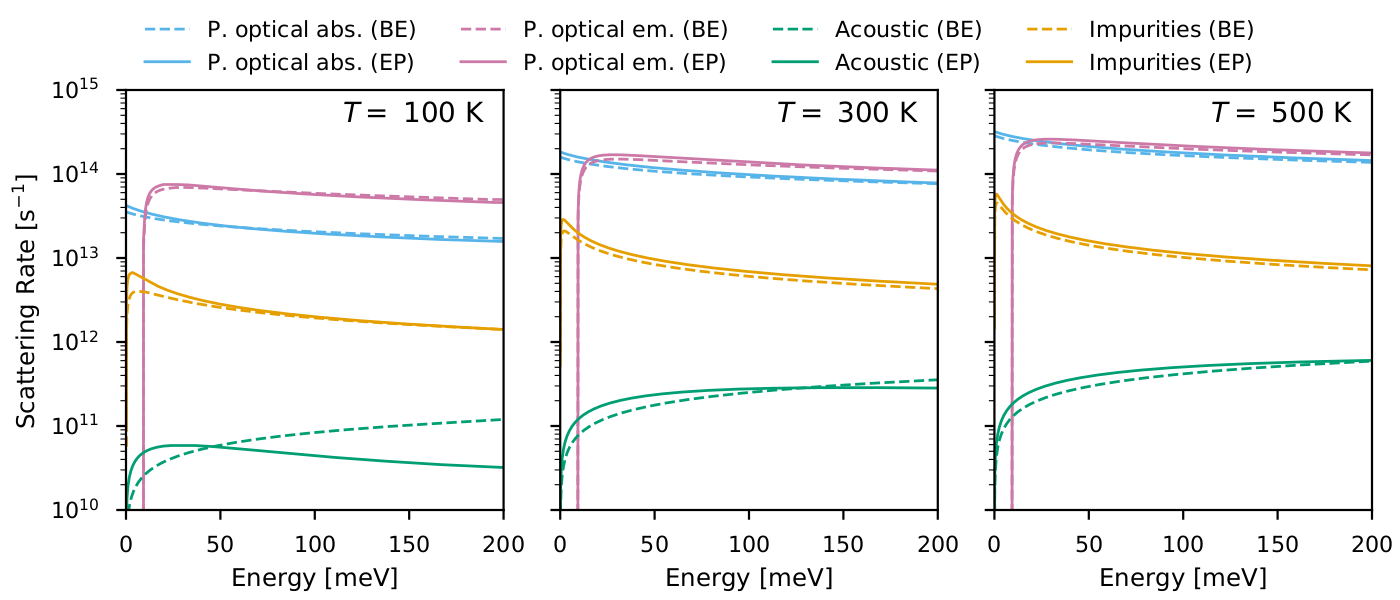}
    \caption{Total scattering rates for band electrons (BE) and electron polarons (EP) in MAPbI$_3$ due to their interaction with polar optical phonons (labelled p. optical abs. and em., for phonon absorption and emission, respectively), acoustic phonons and ionised impurities, at 100, 300 and 500 K. Similar results for MAPbBr$_{3}$ and CsPbI$_{3}$ (as well as the parameters used in the calculations) are presented in the supplementary information.}
    \label{fig:total_rates}
\end{figure}

The \emph{total} scattering rate for a given mechanism, as a function of a particle's wave vector before scattering $\mathbf{K}_{\mathrm{i}}$, is calculated by integrating over all possible final states $\mathbf{K}_{\mathrm{f}}$. Total scattering rates for electron polarons in MAPbI$_{3}$ are plotted in Fig. \ref{fig:total_rates} at 100, 300 and 500 K. To enable comparison with the analogous rates for band electrons, also shown, we plot the rates as functions of the initial kinetic energy, $\hbar^{2}\lvert \mathbf{K}_{\mathrm{i}} \rvert^{2}/2M$. For scattering due to ionised impurities, we assume a density of such defects of $10^{16}$ cm$^{-3}$, which is towards the upper end of the range of values measured in polycrystalline LHP solar cells \cite{leijtens2016carrier, dequilettes2015impact, chen2016extended}.

Of the mechanisms considered, we find that the acoustic phonon scattering rate is the one most significantly affected by polaron formation, while the effects on polar optical and impurity scattering rates are essentially negligible. In all cases, the differences between scattering rates for the the band electron and the electron polaron are reduced as the temperature increases. We also note that polar optical phonon scattering is dominant for both band electrons and electron polarons.

The total scattering rates do not provide a complete picture, however; we must also examine the angular distribution of final states, not least because a strongly anisotropic scattering mechanism will randomise the particle distribution function less rapidly than an isotropic one, all else being equal. After undergoing a scattering event, a particle's \emph{final} wave vector can be fully described by its spherical coordinates in a reference frame defined by its \emph{initial} wave vector. The magnitude of the final wave vector is determined by the delta function in Eqn. \ref{eqn:goldenrule}, and the distribution of the azimuthal angle is constant due to the cylindrical symmetry of the reference frame. Therefore, we need only consider the dependence on the magnitude of the initial wave vector, and the (polar) angle between the final and initial wave vector vectors.

Thus, immediately after undergoing a scattering event due to a given mechanism, the probability per unit solid angle that the particle's current wave vector, $\mathbf{K}_{\mathrm{f}}$ and its wave vector immediately before the scattering event, $\mathbf{K}_{\mathrm{i}}$ are at an angle $\theta$ to each other is

\begin{equation}
    S(\mathbf{K}_{\mathrm{i}},\theta) = \frac{\int \mathrm{d}\lvert \mathbf{K}_{\mathrm{f}} \rvert \lvert \mathbf{K}_{\mathrm{f}} \rvert^{2} S(\mathbf{K}_{\mathrm{i}} \rightarrow \mathbf{K}_{\mathrm{f}})}{\int \mathrm{d}^{3}\mathbf{K}_{\mathrm{f}} S(\mathbf{K}_{\mathrm{i}} \rightarrow \mathbf{K}_{\mathrm{f}})}
\end{equation}

\begin{figure}[t!]
    \centering
    \includegraphics[width=1.0\linewidth]{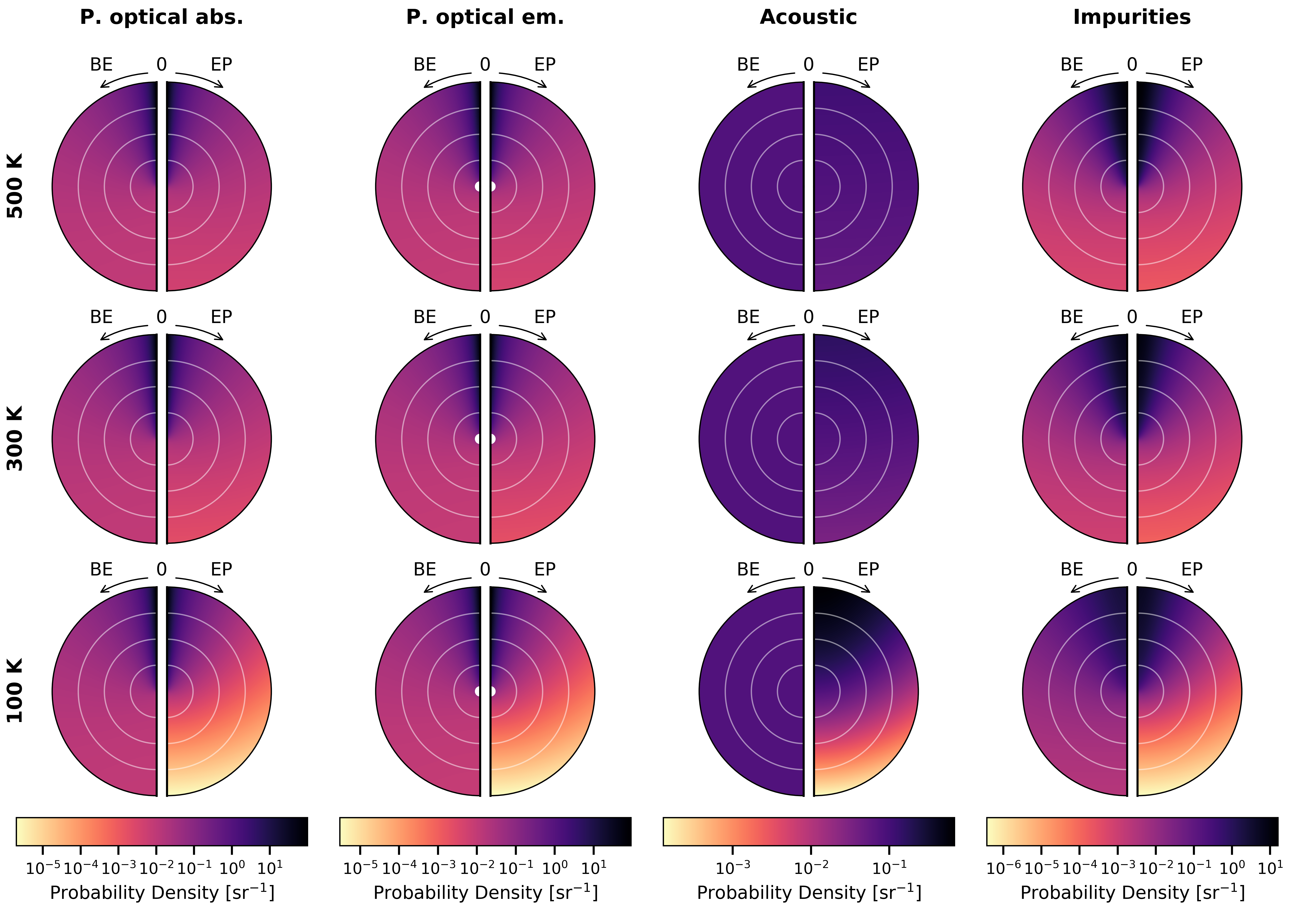}
    \caption{Probability per unit solid angle of particle states after scattering, as a function of the initial particle kinetic energy (radial co-ordinate) and the angle between the initial and final wave vectors (polar co-ordinate). The distributions for band electrons (BE) and electron polarons (EP) are plotted on the left and right halves of each polar plot. The radial circles delineate divisions of 50 meV. Results are presented for MAPbI$_3$ for temperatures of 100, 300 and 500 K; similar results for MAPbBr$_{3}$ and CsPbI$_{3}$ are presented in the supplementary information.
    }
    \label{fig:polar_rates}
\end{figure}

In Fig. \ref{fig:polar_rates}, we plot the distribution for $\theta$ at 300 K; however, we use the initial kinetic energy rather than the magnitude of the initial wave vector as the radial co-ordinate to allow for a more direct comparison to be made with the corresponding distributions for band electrons, which are also plotted.

We see that, as with the total scattering rates, the most strongly affected distribution is that for acoustic phonon scattering, which is isotropic for the band electron case but forward-scattering in the polaron case. However, the impurity and polar optical phonon distributions are already strongly forward scattering for the band electron, and therefore the effects on them of polaron formation are minor. Thus, although the concept of charge carriers being protected from scattering does survive some scrutiny, such effects appear to be non-negligible only in the case of acoustic phonon scattering. Furthermore, we expect their influence to be marginal since polar optical phonons remain the dominant cause of scattering in MAPbI$_{3}$.

In order to investigate more directly the effects of polaron formation on an experimentally measurable quantity, we use the temperature dependent effective masses and scattering rates to calculate the low-field charge carrier mobility of MAPbI$_{3}$ for electron polarons at temperatures in the range $50-500$ K. Such calculations contribute to an ongoing discussion in the literature regarding the apparently anomalous temperature dependence of the mobility in MAPbI$_{3}$. Briefly, conventional semiconductor theory suggests a characteristic temperature dependence of the mobility for each of the three scattering mechanisms considered herein: $T^{-0.5}$ (in the high temperature limit) for optical phonon scattering, $T^{-1.5}$ for acoustic phonon scattering, and $T^{1.5}$ for ionised impurities \cite{jacoboni2010theory}. Therefore, by measuring the temperature dependence of the mobility, one can, in principle, deduce the dominant scattering mechanism in the material. As a highly polar material, one would expect that polar optical phonon scattering should be dominant \cite{herz2018lattice}. However, a $\sim T^{-1.5}$ dependence of the mobility has been measured for MAPbI$_{3}$ using a variety of techniques \cite{milot2015temperature, savenije2014thermally, shrestha2018assessing, biewald2019temperature}. Polaron formation has been invoked as a possible resolution to the apparent discrepancy \cite{zhu2015charge, mante2017electron, zhang2017charge}.

Here, we calculate band-electron and electron-polaron mobilities within the framework of semi-classical Boltzmann transport theory. Such an approach is standard for band-electron transport (see e.g. Refs. \cite{howarth1953theory, blatt1957theory, ashcroft1976solid}), and it is the basis for the so-called ``Kadanoff mobility'' expression for polarons \cite{kadanoff1963boltzmann, kadanoff1964green, osaka1966polaron, osaka1973note, okamoto1974polaron} (see Ref. \cite{peeters1984theory} for a pedagogical overview of the Boltzmann equation for polarons and its relation to numerous commonly applied expressions for polaron mobility). We note that it has previously been asserted that the Boltzmann equation for polarons is strictly valid only at low temperatures, due to the non-negligible probability of excited internal states being occupied at elevated temperatures \cite{kadanoff1963boltzmann, kadanoff1964green, peeters1984theory}; however, as previously discussed, the polarons are overwhelmingly likely to be in the ground state at equilibrium over a wide range of temperatures.

A number of studies of band electron \cite{zhao2016intrinsic, kang2018intrinsic, ponce2019origin} and polaron mobilities \cite{sendner2016optical, frost2017calculating, zhao2017low, whalley2019impact, miyata2017large} at different levels of approximation. Typically, these calculations either make use of the relaxation time approximation, and/or limit the scattering to a single mechanism. Instead, we compute the mobility from the distribution obtained via the ensemble Monte Carlo method \cite{fawcett1970monte, jacoboni1989monte}, which circumvents the need for the relaxation time approximation and also allows us to include all of the scattering mechanisms already presented, namely scattering due to polar optical phonon absorption and emission, acoustic phonons, and ionised impurities. Our simulations were performed using a code that was written in-house; a brief description of the the ensemble Monte Carlo method can be found in the Methods section, along with a full list of material parameters that were used as inputs to the model. We note that all parameters are taken from \textit{ab initio} studies in the literature.

\begin{figure}[h!]
    \centering
    \includegraphics[width=0.5\linewidth]{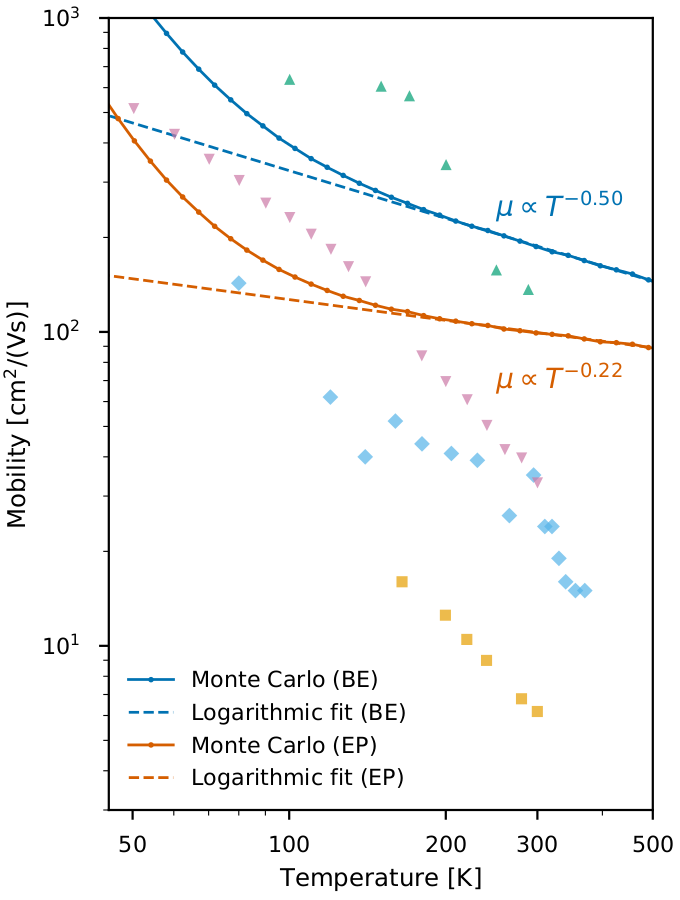}
    \caption{Temperature dependence of the mobility of band electrons (BE) and electron polarons (EP) in MAPbI$_{3}$, as calculated using the ensemble Monte Carlo method. Straight lines representing simple power laws are fitted to values above 200 K. Experimental data from the literature are also shown: blue diamonds from Ref. \cite{milot2015temperature}, yellow squares from Ref. \cite{savenije2014thermally}, green upward triangles from Ref. \cite{shrestha2018assessing} and pink downward triangles from Ref. \cite{biewald2019temperature}. Similar results for MAPbBr$_{3}$ and CsPbI$_{3}$ are presented in the supplementary information.} 
    \label{fig:mobilities}
\end{figure}

The results of the simulations for electrons in MAPbI$_{3}$ are shown in Fig. \ref{fig:mobilities}. We also plot data from several experimental studies of charge carrier mobility in MAPbI$_{3}$, in order to highlight the wide range of values and the approximate agreement of the temperature dependence of $\sim T^{-1.5}$ \cite{savenije2014thermally, milot2015temperature, shrestha2018assessing, biewald2019temperature}. The large spread in data is accounted for by the variety of measurement techniques and material morphologies; measurement techniques range from photoconductivity to time-of-flight methods, with some measurements being unable to distinguish between electron and hole transport, and we also expect the varying degrees of crystallinity to affect the values measured.


Conventional Fr\"ohlich scattering theory of band electrons predicts a temperature dependence of $T^{-0.5}$ at high temperatures \cite{petritz1955mobility}. Fitting a simple power law to our theoretical data points for temperatures above 200 K recovers this result to at least two decimal places. The electron--polaron mobility has been calculated using the same method, with the only alterations being to the particle mass and scattering rates. Figure \ref{fig:mobilities} shows that the most significant change in charge carrier mobility due to polaronic effects is its reduction by a factor of almost 2 at room temperature. At increased temperatures we find that the band and polaron results begin to converge as the differences in carrier masses and scattering rates reduce. The temperature dependence of the mobility extracted for values above 200 K gives $T^{-0.22}$. This is shallower than the band electron result, and further removed from the experimentally observed $\sim T^{-1.5}$ dependence. Therefore, we conclude that polaronic effects play a relatively minor part in charge carrier dynamics here, and do not provide an explanation for the unusual temperature dependence of charge carrier mobility in MAPbI$_{3}$.


In conclusion, we have presented an analysis of the effects of polaron formation in MAPbI$_{3}$ on carrier scattering and mobilities. The results for MAPbBr$_{3}$ and CsPbI$_{3}$, which are given in the supplementary information, are qualitatively similar to those already presented for MAPbI$_{3}$, indicating that our conclusions apply quite generally to LHPs.

By minimising the polaron free energy due to Osaka, we determined the temperature dependence of the polaron mass, oscillator frequency and ground state occupancy. We found that polarons in MAPbI$_{3}$ are almost all in their ground state at equilibrium, and we assumed, close to equilibrium. We derived rates for polaron scattering due to acoustic phonons and ionised impurities, and computed the rates, along with those for polar optical phonon absorption and emission. We found that scattering of electron polarons due to acoustic phonons is the most significantly different from that of band electrons, with the most striking effect being the change from an isotropic final state distribution for band electrons, to an anisotropic one for polarons. In constrast, the rates and final state distributions for scattering by polar optical phonons and ionised impurities are relatively little affected by polaron formation.

The ensemble Monte Carlo method was used to compute distribution functions satisfying Boltzmann transport equations for electron polarons and band electrons, from which we derived temperature dependent mobilities. We found that the polaron mobility is always lower than that of band electrons, with a smaller exponent for the temperature dependence, and therefore we believe that other possible explanations for the $T^{-1.5}$ dependence of mobility in MAPbI$_{3}$ observed in experiments must be considered. 

Looking beyond the halide perovskites, our results invite the question of when polaronic effects might be expected to be significant. On the one hand, we expect that strong coupling to polar optical phonons should be a criterion since it determines the polaron's mass, yet scattering by those phonons is little affected by polaron formation. In particular, based on our results, it would be fruitful to identify a regime in which acoustic phonon scattering plays a more significant role, since the final state distribution is more strongly altered by polaron formation.

Finally, we note that the computational framework used in this study is very general, and can be readily extended to include other scattering mechanisms and spatial inhomogeneity, such as that encountered in optoelectronic devices, and be used to examine non-equilibrium polaron dynamics beyond computations of the low field mobility, where analytical expressions become increasingly difficult to obtain.

\section*{Methods}

The ensemble Monte Carlo method yields particle distribution functions satisfying the Boltzmann transport equation (BTE) by simulating trajectories of an ensemble of particles \cite{fawcett1970monte}. The motion of each particle follows the same assumptions made in the derivation of the BTE: particles undergo periods of free flight under the influence of an applied field, which are interrupted by instantaneous scattering events that change their momenta. Scattering events are chosen by sampling from the Poisson distributed total rate functions using a random number generator; following a selection of a scattering mechanism, the final momentum vector of the particle is then chosen by sampling the distributions of polar and azimuthal scattering angles. More detailed descriptions of the details of computational implementation can be found in Refs. \cite{hockney1988computer, tomizawa1993numerical}.

To produce the data reported herein, we simulated trajectories for $10^6$ particles under an applied field of 1 kV cm$^{-1}$ for a total time of 100 ps. The mean particle velocity in the direction of the applied field was recorded every 0.1 ps after an initial simulation period of 10 ps, in order to reach steady state. Scattering rates were pre-tabulated as a function of initial particle energy at intervals of $2 \times 10^{-4}$ eV for scattering due to polar optical phonons (emission and absorption), acoustic phonons and ionised impurities. Convergence of the low-field mobility was confirmed by performing calculations at lower fields, which yielded identical results. The material parameters used in the simulations are collected in Table \ref{tab:matparams}.

\begin{table}[h]
    \centering
    \caption{Material parameters for the three lead--halide perovskites, MAPbI$_{3}$, MAPbBr$_{3}$ and CsPbI$_{3}$, that were included in our study. Parameters were chosen to align with several previous studies, and sourced entirely from \textit{ab initio} calculations. The elastic constants $c_{\rm L}$ were calculated from the mean value of $C_{11}$, $C_{22}$ and $C_{33}$ (elastic constants from Ref. \cite{lee2018competition}).}
    \label{tab:matparams}
    \begin{tabularx}{\textwidth}{ X | X X X }
        \hline
        Parameter (units) & MAPbI$_{3}$ & MAPbBr$_{3}$ & CsPbI$_{3}$ \\
        \hline
        $m$ ($m_{\rm e}$)           & 0.15 \cite{brivio2014relativistic} & 0.27 \cite{mosconi2015electronic} & 0.17 \cite{ponce2019origin}      \\
        $c_{\rm L}$ (GPa)           & 32 \cite{lee2018competition}       & 37   \cite{lee2018competition}    & 27   \cite{lee2018competition}   \\
        $\Xi_{\rm d}$ (eV)          & -2.13 \cite{lee2018competition}    & -2.67 \cite{lee2018competition}   & -2.13 \cite{lee2018competition}  \\
        $\omega_{\rm 0}/2\pi$ (THz) & 2.25 \cite{frost2017calculating}   & 3.46 \cite{zhao2017low}           & 2.57 \cite{frost2017calculating} \\
        $\varepsilon_{\rm 0}$       & 25.7 \cite{brivio2014relativistic} & 25 \cite{zhao2017low}             & 18.1 \cite{frost2017calculating} \\
        $\varepsilon_{\infty}$      & 4.5  \cite{brivio2013structural}   & 6.7 \cite{zhao2017low}            & 6.1 \cite{frost2017calculating}  \\
        \hline
    \end{tabularx}
\end{table}

All figures were produced using the Python graphical environment matplotlib \cite{hunter2007matplotlib}.

\section*{Acknowledgements}

This work was funded by the United Kingdom's Engineering and Physical Sciences Research Council (EPSRC) through the Supersolar hub project and the Centre for Doctoral Training in New and Sustainable Photovoltaics (EP/L01551X). It also received funding from the  European Union's Horizon 2020 research and innovation programme under the EoCoE II project (824158). We thank Prof. L. M. Peter for reading and commenting on the manuscript.

\section*{Author Contributions}

MJW and ABW initiated and designed the study. MJW derived the polaron scattering rates. MJW and LADI wrote the ensemble Monte Carlo code and performed the associated simulations. MJW prepared the manuscript with contributions from LADI and ABW.

\end{document}